\documentclass[aps,preprint,amsmath,amssymb]{revtex4}
\usepackage{graphicx}
\newcommand{\nn}{\nonumber}
\newcommand{\be}{\begin{eqnarray}}
\newcommand{\ee}{\end{eqnarray}}

\begin{document}

\title{\Large \bf Study of $B \to K_0^* (1430)\ell\bar{\ell} $ Decays}

\author{ \bf \large Chuan-Hung Chen$^{1,2}$, Chao-Qiang Geng$^{3,4}$,
 Chong-Chung Lih$^{5}$, and Chun-Chu Liu$^{6}$
 }

\affiliation{  $^{1}$Department of Physics, National Cheng-Kung
University, Tainan 701, Taiwan \\
$^{2}$National Center for Theoretical Sciences, Taiwan\\
$^{3}$Department of Physics, National Tsing-Hua University,
Hsinchu
300, Taiwan  \\
$^{4}$Theory Group, TRIUMF, 4004 Wesbrook Mall, Vancouver, B.C V6T
2A3, Canada\\
$^5$General Education Center, Tzu-Chi  College of Technology
Hualien 970, Taiwan \\
$^6$Department of Electrophysics, National Chiao-Tung University
 Hsinchu 300, Taiwan
 }

\date{\today}

\begin{abstract}
We examine the exclusive rare decays
of $B \to K_{0}^*(1430)\ell\bar{\ell} $ with $K_{0}^*(1430)$ being
the p-wave scalar meson and $\ell=\nu,e,\mu,\tau$
in the standard model.
The form factors  for the $B\to K_{0}^{*}$ transition matrix elements
are evaluated in the light front quark model.
For the decays of
$B \to K_{0}^*\ell\bar{\ell}$, the branching ratios are found to be
$(11.6,1.63,1.62,0.029)\times 10^{-7}$ with $\ell=(\nu,e,\mu,\tau)$
and
the integrated longitudinal lepton polarization asymmetries
 $(-0.97,-0.95,-0.03)$ with $\ell=(e,\mu,\tau)$, respectively.

\end{abstract}

\maketitle %

\section{Introduction}
The suppressed inclusive flavor-changing neutral current (FCNC)
process of $B\to X_s \ell^+ \ell^-$, induced by electroweak
penguin and box diagrams in the standard model (SM), has been
 observed by both BaBar \cite{BaBar1} and Belle
\cite{Belle1}
 with the branching ratio (BR) of $(4.5\pm1.0)\times 10^{-6}$ \cite{pdg}
 for di-lepton masses greater than $0.2\ GeV$,
 where $\ell$ is either an electron or a
muon and $X_s$ is a hadronic recoil system that contains a kaon.
The exclusive decays of $B\to K \ell^+ \ell^-$ and
$B\to K^{*}(892) \ell^+ \ell^-$
have also been
measured with the BRs \cite{pdg} of
$(0.54\pm0.08)\times 10^{-6}$
and $(1.05\pm0.20)\times
10^{-6}$ \cite{BaBar2,Belle2},
which
agree with the theoretically estimated values
 \cite{Ali,CK,MNS,CG-PRD}, respectively.

There have been many investigations of rare $B$ semileptonic decays of
 induced by the
FCNC transition of $b \to s$
\cite{Bdecays} since the CLEO observation \cite{cleo} of $b\to
s\gamma $.
 The studies are even more complete if similar studies for
the p-wave mesons of $B$ decays
such as $B \to K_{0,2}^* (1430)\ell\bar{\ell} $ and
$B \to K_{1A,1B}\ell\bar{\ell} $
 are also included.
  In fact,  the study of $B \to K_{2}^* (1430)\ell^{+}\bar{\ell}^{-} $
has been done
in Ref.
\cite{PRD74-K2}.
It is clear that these FCNC rare
decays are important for not only testing the SM but probing new
physics.
In this report, we concentrate on the exclusive rare decays
of $B \to K_{0}^*\ell\bar{\ell} $, where $K_{0}^*$ represents
the p-wave scalar meson of $K_{0}^* (1430)$ and $\ell$
stands for a charged lepton or neutrino. To obtain the decay rates
and branching ratios, we need to calculate the transition form
factors of $B\to K_{0}^{*}$ due to
the  axial-vector and axial-tensor currents,
respectively, in the standard model.
We will use the framework of the light front
quark model (LFQM) \cite{hwcw,tensor,Cheng1} to evaluate  these form
factors.

This report is organized as follows:  We present
the relevant formulas in Sec. II. First, we  give
the effective Hamiltonians
for $B \to K_0^* \ell\bar{\ell}$
 induced by $b\to s\ell \,\bar{\ell}$.
Then, we calculate the
hadronic form factors for the $B\to K_{0}^{*}$
transition in the LFQM.
Finally, we study
the branching ratios and  polarization asymmetries of the  decays.
In Sec. III, we show
our numerical results on form factors and the
physical quantities of the  decays.
We give our conclusions in Sec. IV.

\section{The formulas}
To study the exclusive decays of
$B \to K_0^* \ell\bar{\ell}$,
we start with the effective
Hamiltonians at the quark level,
 given by
  \be
    {\cal H}(b \to s\nu \bar{\nu})&=&\frac{G_{F}}{\sqrt{2}} \frac{\alpha }{2\pi \sin^{2}\theta _{W}}
     \lambda _{t} D\left( x_{t}\right) \bar{b}\gamma _{\mu }\left( 1-\gamma
      _{5}\right) s\bar{\nu_{\ell} }\gamma _{\mu}\left( 1-\gamma _{5}\right)
     \nu_{\ell} \,,
 \nonumber\\
{\cal H}(b \to s \ell^+ \ell^-)&=&\frac{G_F\alpha \lambda _t}{\sqrt{2}\pi }\left[ C_9^{eff}
       ( m_b ) \bar{s}_L\gamma _{\mu }b_L\ \bar{\ell}
       \gamma ^{\mu }\ell +C_{10}\bar{s} _L\gamma _{\mu }b_L\
       \bar{\ell}\gamma ^{\mu }\gamma _5 \ell
       ~~~~~~~~~~~~~~\right.\nn \\
    &&~~~~~~~~~~~~\left.-\frac{ 2m_b C_7(m_b) }{q^2}
       \bar{s}_L i\sigma _{\mu \nu }
       q^{\nu }b_R\ \bar{\ell}\gamma ^{\mu }\ell\right]\,,
       \label{Hll}
  \ee
where $x_t \equiv m_t^2 /m_W^2$, $\lambda _{t}=V_{ts}^*V_{tb}$,
 $D(x_t)$ is the top-quark loop function \cite{geng1,Buras0}
 and $C_{i}$ are the Wilson
coefficients (WCs) with their explicit expressions given in
Ref.~\cite{BBL}.
 In particular,  $C^{\rm eff}_{9}$,
  which contains the contribution from the on-shell charm-loop,  is given by
 \cite{BBL}
\begin{eqnarray}
C_{9}^{\rm eff}(\mu)&=&C_{9}( \mu ) +\left( 3C_{1}\left( \mu \right)
+C_{2}\left( \mu \right) \right)  h\left( z,\hat{s}\right)
\,, \nonumber \\
h(z,\hat{s})&=&-\frac{8}{9}\ln\frac{m_b}{\mu}-\frac{8}{9}\ln z
+\frac{8}{27} +\frac{4}{9}x  -\frac{2}{9}(2+x)|1-x|^{1/2} \nonumber
\\
&\times& \left\{
  \begin{array}{c}
    \ln \left|\frac{\sqrt{1-x}+1}{\sqrt{1-x}-1} \right|-i\, \pi, \  {\rm for}\ x\equiv 4z^2/\hat{s}<1 \, , \\
    2\, arctan\frac{1}{\sqrt{x-1}},\   {\rm for}\ x\equiv 4z^2/\hat{s}>1   \, ,\\
  \end{array}
\right.
\label{C9eff}
\end{eqnarray}
where
$z=m_c/m_b$ and $\hat{s}=q^2/m^2_b$
with $q^2$ being the invariant mass of the dilepton.
Here, we have ignored the resonant
contributions \cite{Res,CGPRD66} as 
the modes such as $B\to J/\Psi K^{*}_{0}$ have not been seen yet.
To calculate the decay rates, we need to evaluate the hadronic
matrix elements for the $B\to K^{*}_{0}$ transition involving
the axial-vector  and axial-tensor currents, whereas
those from the vector  and tensor  ones are zero due
to the parity conservation in strong interactions. In the following,
we will study the matrix elements in the
%
 LFQM \cite{Ter,Chung}, which has been successfully
 applied to many weak processes
with
  the heavy-to-heavy and
heavy-to-light transitions in the timelike regions
\cite{hwcw,tensor,Cheng1,time1}.

In the LFQM,
a meson bound state consisting of a heavy quark $q_1$ and an
antiquark $q_2$ with the total momentum $P$ and spin $S$ can be
written as \be
        |B(P, S, S_z)\rangle
              &=&\int  {dp_1^+d^2p_{1\bot}\over 2(2\pi)^3}
                       {dp_2^+d^2p_{2\bot}\over 2(2\pi)^3}
                ~2(2\pi)^3 \delta^3(\tilde
                P-\tilde p_1-\tilde p_2)~\nn\\
        &&\times \sum_{\lambda_1,\lambda_2}
                \Psi^{SS_z}(\tilde p_1,\tilde p_2,\lambda_1,\lambda_2)~
                |q_1(p_1,\lambda_1) \bar q_2(p_2,\lambda_2)\rangle,
\label{bound1} \ee where $p_1$ and $p_2$ are the on-mass-shell
light front momenta
 \be
        \tilde p=(p^+, p_\bot)~, \quad p_\bot = (p^1, p^2)~,
                \quad p^- = {m^2+p_\bot^2\over p^+},
\ee with \be
        && p^+_1=(1-x) P^+, \quad p^+_2=x P^+, \nn \\
        && p_{1\bot}=(1-x) P_\bot+k_\bot, \quad p_{2\bot}=x
        P_\bot-k_\bot\,.
\ee Here $(x,k_\perp)$ are the light-front relative momentum
variables and $\vec{k}_\perp$ is the component of the internal
momentum $\vec{k}=(\vec{k}_\perp,k_z)$. The momentum-space
wave-function $\Psi^{SS_z}$ in Eq. (\ref{bound1}) can be expressed
as
   \be
        \Psi^{SS_z}(\tilde p_1,\tilde p_2,\lambda_1,\lambda_2)
                = R^{SS_z}_{\lambda_1\lambda_2}(x,k_\bot)~ \phi^{(p)}(x, k_\bot),
        \label{momentumspace}
   \ee
where $R^{SS_z}_{\lambda_1\lambda_2}$ constructs a state of
a definite spin ($S,S_z$) out of the light-front helicity
 ($\lambda_1,\lambda_2$) eigenstates and $\phi^{(p)}(x,k_\bot)$
describes the momentum distribution of the constituents in the
bound state for the s-wave (p-wave) meson. Explicitly, it is more
convenient to use the covariant form for
$R^{SS_z}_{\lambda_1\lambda_2}$ \cite{Cheng1}, given by

   \be
        R^{SS_z}_{\lambda_1\lambda_2}(x,k_\bot)
                =h_M
        ~\bar u(p_1,\lambda_1)\Gamma_{M} v(p_2,\lambda_2), \
      (M=P_i,S_f)
        \label{covariant}
  \ee
where
  \be
    \Gamma_{P_i}&=&\gamma_{5}\,,\ \
 h_{P_i}=(m_{P_i}^2-M_{0}^2) \sqrt{\frac{x(1-x)}{N_c}}\frac{1}{\sqrt{2}
        {\widetilde M_{0}}}\,, \nn \\
  \Gamma_{S_f}&=&-i\,,\  \
h_{S_f}=(m_{S_f}^2-M_{0}^2)
\sqrt{\frac{x(1-x)}{N_c}}\frac{1}{\sqrt{2}
        {\widetilde M_{0}}} \frac{{\widetilde M_{0}}^2}
        {\sqrt{3}M_{0}}  \,,
  \ee
 for the initial s-wave pseudoscalar $(P_i)$ and final p-wave scalar
$(S_f)$ mesons of $B$ and $K_0^*$, respectively, with
  \be
    {\widetilde M_0} &\equiv &\sqrt{M_0^2-(m_{1}-m_{2})^2}
     \label{M0}\,,\nn\\
        M_0^2&=&{ m_{1}^2+k_\bot^2\over (1-x)}+{ m_{2}^2+k_\bot^2\over
        x}\,.
  \ee
In our calculations, we use the Gaussian type wave functions
  \be\label{wave}
   \phi(x, k_\bot)&=&4\left(\frac{\pi}{\omega_M ^2}\right)^{3/4}
   \sqrt{dk_z \over dx} \exp\left( -
   \frac{\vec{k}^2}{2\omega_M ^2}\right)\,,  \nn\\
   \phi^p(x, k_\bot)&=& \sqrt{\frac{2}{\omega_M^2}}~ \phi(x, k_\bot) \,,
  \ee
  for
the s-wave and p-wave mesons,
respectively, where $\omega _M$ is the meson scale parameter and
$k_z$ is defined through
  \be
    1-x=\frac{e_1-k_z}{e_1+e_2}\,,\quad  x=\frac{e_2+k_z}{e_1+e_2}
  \ee
with $e_i=\sqrt{m^2_{i}+\vec{k}^2}$. Then, we have
   \be
       M_0=e_1+e_2\,, \quad
       k_z=\frac{xM_0}{2}-\frac{m^2_{2}+k^2_{\perp}}{2xM_0}\,,
   \ee
and
  \be
     \frac{dk_z}{dx}=\frac{e_1e_2}{x(1-x)M_0}\,.
  \ee
 We normalize the meson state as
\be
        \langle B(P',S',S'_z)|B(P,S,S_z)\rangle = 2(2\pi)^3 P^+
        \delta^3(\tilde P'- \tilde P)\delta_{S'S}\delta_{S'_zS_z}~,
\label{waveno}
 \ee
so that the normalization condition of the momentum distribution
function can be obtained by
  \be
       \int {dx\,d^2k_\bot\over 2(2\pi)^3}~|\phi(x,k_\bot)|^2 = 1.
    \label{momnor}
  \ee

 We are now ready to calculate the matrix elements of the
 $P_{i} \to S_f$ transition, which can be defined by
 \be
  \left\langle S_f(p_f)|\ A_\mu \ |P_i(p_i)\right\rangle
      &=& -i [u_+(q^2)P_{\mu} + u_-(q^2)q_{\mu}]\,,
    \nn\\
  \left\langle S_f(p_f) |\ T^5_{\mu\nu} q^\nu \ |P_i(p_i)
          \right\rangle &=& \frac{-i}{m_{P_i}+m_{S_f}}\left[ q^2P_{\mu }-\left( P\cdot
            q\right) q_{\mu } \right] F_T (q^2) \,,   \label{matrixPS}
 \ee
where $A_{\mu}=\bar{q}_f \gamma _{\mu} \gamma _5 q_i $,
$T^5_{\mu}=\bar{q}_f i \sigma _{\mu \nu} \gamma _{5} q_i $,
$P=p_i+p_f$ and $q=p_i-p_f$
with
the initial (final) meson bound state
$q_i\bar{q}_3$  ($q_f\bar{q}_3$).
We note that all  form factors
will be studied in the timelike physical
meson decay region of $0\leq q^2 \leq (m_{P_i}-m_{S_f})^2$.
The form factors
in Eq. (\ref{matrixPS}) are found to be
  \be
     u_+ (q^2)&=& \frac{(1-r_-)H(r_+)-(1-r_+)H(r_-)}{r_+-r_-} \,,
     \nn\\
     u_- (q^2)&=& \frac{(1+r_-)H(r_+)-(1+r_+)H(r_-)}{r_+-r_-} \,,
\nn\\
F_T(q^2) &=& - \int_0^rdx \int \frac{d^2k_{\perp}}{2(2\pi)^3}
                \frac{{\widetilde M_0}^{~2}}{2\sqrt{3}M_0}\
                \phi_{S_f}^{p*}(x',k_{\perp})\phi_{P_i}(x,k_{\perp})
                 \nn \\
                &&\frac{m_{P_i}+m_{S_f}}{(1+2r)q^2-(m^2_{P_i}-m^2_{S_f})}
                \frac{A}{\sqrt{{\cal A}^2_{P_i}+k^2_{\perp}}
                 \sqrt{{\cal A}^2_{S_f}+  k^2_{\perp}}} \,,
  \ee
where
$r\equiv p_f^+/p_i^+$,
$x'=x/r$,
  \be
   r_{\pm} &=& \frac{m_{S_f}}{m_{P_i}}[v_i\cdot v_f \pm
        \sqrt{(v_i \cdot v_f)^2-1}]\,, ~~
   \left( v_i\cdot v_f =\frac{m^2_{P_i}+m^2_{S_f}-q^2}
           {2m_{P_i}m_{S_f}}\right)
\nn\\
 H(r) &=& -\int_0^rdx \int \frac{d^2k_{\perp}}{2(2\pi)^3}
        \frac{{\widetilde M_0}^{~2}}{2\sqrt{3}M_0}
        \phi_{S_f}^{p*}(x',k_{\perp}) \phi_{P_i}(x,k_{\perp}) \nn \\
        &&
        \frac{[m_{q_i}x+m_{q_3}(1-x)][-m_{q_f}x'+m_{q_3}(1-x')]+k^2_{\perp}}
        {\sqrt{{\cal A}^2_{P_i}+k^2_{\perp}}\sqrt{{\cal A}^2_{S_f}+
        k^2_{\perp}}} \,,
 %
\nn\\
    A &=& \frac{1}{\sqrt{xx'}(1-x)(1-x')} \bigg\{\left[ x m_{q_i}+(1-x)
         m_{q_3} \right] \left[ -x' m_{q_f}+(1-x')m_{q_3}\right] \nn\\
     &&\left[x(1-x')m_{q_i}-x'(1-x)m_{q_3}\right]+k^{2}_{\perp}
       \left[-x'(1-x')(2x-1)m_{q_f} \right. \nn \\
     &&\left. +(x-x')(x+x'-2xx')m_{q_3}+x(1-x)(1-2x')m_{q_i}
     \right]\bigg\}\,, \nn \\
      {\cal A}_{P_{i}(S_{f})}&=&m_{q_{i}(q_f)}x^{(')}+m_{q_3}(1-x^{(')})\,.
  \ee
The sign $+(-)$ of $r_{+(-)}$ represents the final meson
recoiling in the positive (negative) z-direction relative to the
initial meson.

We note that to evaluate the form factors, we have to fix the meson scale
parameters $\omega_{B}$ and $\omega_{K^{*}_{0}}$ in the meson
wave functions in Eq. (\ref{wave}) by some known parameters such
as the meson decay constants, defined by
       \be
      f_{P_{i}} &=& \sqrt{24} \int {dx\,d^2k_\perp\over 2(2\pi)^3}\,
        \phi(x, k_\perp) \,\frac{m_{q_i}x+m_{q_3}(1-x)}{
        \sqrt{{{\cal A}_{P_i}}^2+k_\perp^2}}, 
        \nonumber \\
 f_{S_f} &=& \sqrt{24} \int {dx'\,d^2k_\perp\over 2(2\pi)^3}\,
        \frac{{\widetilde M_0}^{~2}}{2\sqrt{3}M_0}\phi^p(x', k_\perp) \,\frac{m_{q_1}x'-m_{q_2}(1-x')}{
        \sqrt{{\cal A}^2+k_\perp^2}},
%
   \ee     

 By using Eqs. (\ref{Hll}) and (\ref{matrixPS}), we derive the
differential decay rates of $B\to K^{*}_{0}
\ell\bar{\ell}$ as
  \be
    \frac{d\Gamma \left(  B\rightarrow K^{*}_{0}\nu
           \bar{\nu }\right) }{ds}
       &=&\frac{G_F^2|\lambda _t|^2
         \alpha _{em}^2|D\left( x_{t}\right) |^2m_{B}^5}{2^8\pi ^5
         \sin^4\theta_W}~\varphi_{ K^{*}_{0} }^{1/2}~|u_+| ^2 \,,
\nn\\
    \frac{d\Gamma \left( B\to K^{*}_{0} \ell^+\ell^-\right) }{ds}
        &=&\frac{G_F^2|\lambda _t|^2m_{B}^5\alpha _{em}^2}
        {3\cdot 2^9\pi ^5}~(1-\frac{4t}{s})^{\frac{1}{2}}~\varphi_{K^{*}_{0}} ^{1/2}
        \left[ \left(1+\frac{2t}
        {s}\right) \alpha_{K^{*}_{0}} +t~\delta_{K^{*}_{0}} \right] \,,
      \label{Rate3}
\ee
where
 \be
    s&=&q^{2}/m_{B}^{2}\,, \quad t=m_{l}^{2}/m_{B}^{2}\,,\quad
      r_{K^{*}_{0}}=m_{K^{*}_{0}}^{2}/m_{B}^{2}\,, \quad
     \nonumber \\
    \varphi_{K^{*}_{0}}&=&\left( 1-r_{K^{*}_{0}}\right) ^{2}-2s\left( 1+r_{K^{*}_{0}}\right)
      +s^{2}\,,
\nn\\
    \alpha_{K^{*}_{0}} &=& \varphi_{K^{*}_{0}} \left( |C_9^{eff}u_+
       -\frac{2C_7 F_T}{1+\sqrt{r_{K^{*}_{0}}}}|^2
       +|C_{10}u_+|^2\right) \,, \nonumber \\
    \delta_{K^{*}_{0}} &=& 6|C_{10}|^2\{\left[2\left(1+r_{K^{*}_{0}}\right)
                  -s\right] |u_+|^2+2\left( 1-r_{K^{*}_{0}}\right)
                 Re(u_+u_-^*)+s|u_-|^2\}~\,.
  \ee
When the polarization of the charged lepton is in the longitudinal direction, i.e.
$\hat{n}=\mathbf{e}_L= \vec{p}_\ell/ |\vec{p}_\ell|=\pm1$,
we can also define the longitudinal lepton polarization asymmetry in
 $B\to K^{*}_{0}\ell^+\ell^-$ as follows \cite{GB,GK}
\be
   P_L(s)=\frac{~~{d\Gamma(\hat{n}=-1)\over ds}~-~
                  {d\Gamma(\hat{n}=1)\over ds}~~}
               {~~{d\Gamma(\hat{n}=-1)\over ds}~+~
                  {d\Gamma(\hat{n}=1)\over ds}~~}\,.
\label{PL}
 \ee
 From Eq. (\ref{PL}), we find that
 \be
     P_L &=& \frac{2\left( 1-\frac{4t}{s}\right)
     ^{1\over2}} {\left(1+\frac{2t}{s}\right)\alpha_{K^{*}_{0}}
     +t ~\delta_{K^{*}_{0}}}
     ~Re~\left[\varphi_{K^{*}_{0}}\left(C_9^{eff}u_+
     -2\frac{C_7F_T}{1+\sqrt{r_{K^{*}_{0}}}}
     \right)(C_{10}u_+)^*\right]\,
 \ee
 in $B\to K^{*}_{0}\ell^+\ell^- $.
 We remark that there is no forward-backward asymmetry for
 $B \to K^*_0\ell^{+}\ell^{-}$ in the SM similar to other pseudoscalar to pseudoscalar dilepton decays such as
$P_{i} \to P_{f}\ell^{+}\ell^{-}$  with $P_{i}=K(B)$
and $P_{f}=\pi (K)$ \cite{GengHo}.

\section{Numerical results}

\begin{figure}[h]\vskip 2cm
 \includegraphics{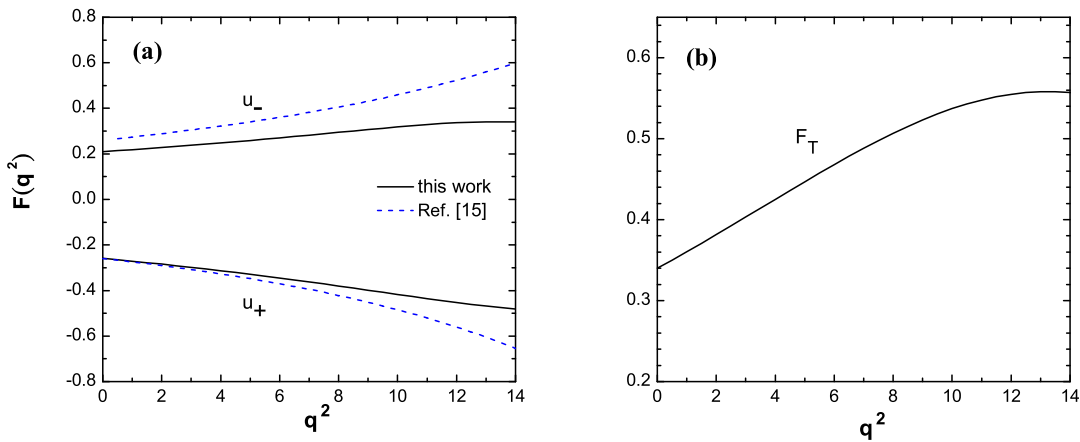} \vskip 5.5cm
 \caption{Form Factors as functions
of $q^{2}$ for (a) $u_+(q^2)$ and $u_-(q^2)$ and (b) $F_T(q^2)$}
  \label{formfactors}
\end{figure}

\begin{figure}[h]\vskip 0cm
 \includegraphics{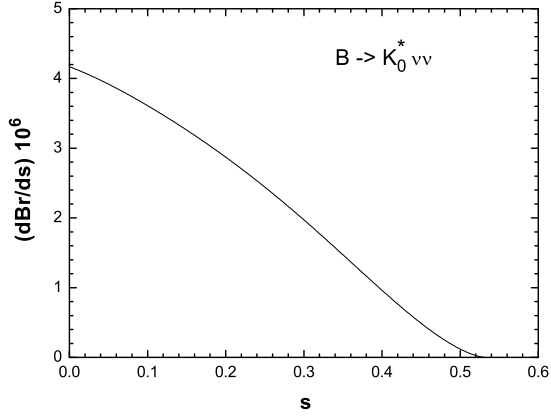} \vskip 7cm
 \caption{Differential decay branching ratio for
 $B\to K_0^* \nu \bar{\nu}$ as a function of $s=q^2/m^2_{B}$.
 }
  \label{Figbrnn}
\end{figure}

\begin{figure}[h]
\vskip 0cm
 \includegraphics{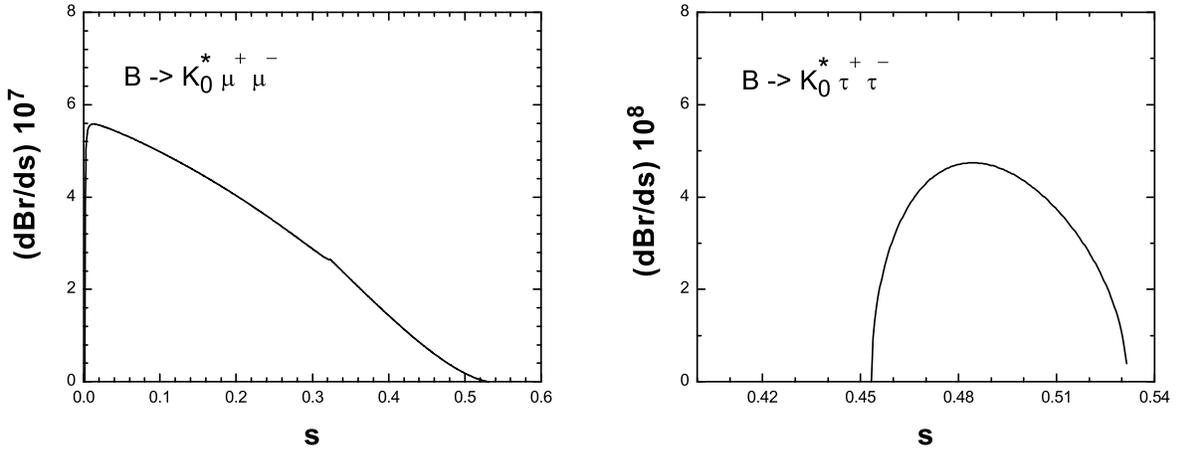}
         \vskip 7cm
 \caption{Differential decay branching ratios for
  $B \to  K^*_0 \mu ^{+}\mu ^{-}$ and
  $B \to  K^*_0 \tau ^{+}\tau ^{-}$ as  functions of $s=q^2/m^2_{B}$.
 }
  \label{Figbrll}
\end{figure}

\begin{figure}[h]\vskip 0cm
 \includegraphics{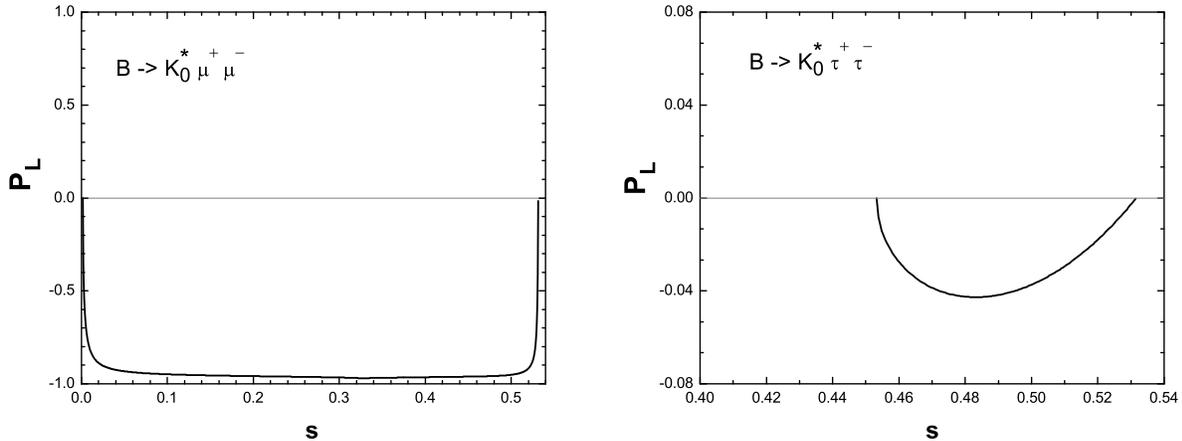}
          \vskip 7cm
 \caption{Longitudinal lepton polarization
asymmetries for $B\to K_0^* \ell^+\ell^-$.
}
  \label{PLbrll}
\end{figure}

In our numerical study of
the hadronic matrix elements for the $B \to K^*_0$
transition, 
we fix the quark masses to be
$m_{b}=4.64$, $m_{s}=0.37$ and $m_{u,d}=0.26$ GeV and use
the meson decay constants to determine the meson scale
 parameters as shown in Table \ref{para}.
\begin{table}[h]
   \caption{Meson decay constants and scale parameters
   (in units of GeV).}
  \label{para}
   \begin{center}
  \begin{tabular}{|c|c||c|c|}
   \hline
   $f_{B}$ & $\omega_{B}$ & $f_{K^{*}_{0}}$& $\omega_{K^{*}_{0}}$\\
   \hline
   0.16 & 0.4763 & 0.015 & 0.2106\\ \hline
   0.18 & 0.5239 & 0.021 & 0.3001\\ \hline
   0.20 & 0.5713 & 0.025 & 0.3837 \\ \hline
%
   \end{tabular}
   \end{center}
\end{table}
We note that the current direct measurement of $f_{B}$ is
$0.229^{+0.036+0.034}_{-0.031-0.037}$ GeV \cite{BellefB}, whereas
there is no experimental information on $f_{K^{*}_{0}}$.
Since  $f_{B}$ and $f_{K^{*}_{0}}$ are
  fixed to be $0.18$ and $0.021$ GeV in Ref. \cite{Cheng1}, respectively,
  we will also use these values in  our numerical analysis and briefly discuss
  other values at the end.
  Our results for the form factors at $q^{2}=0$
are given in Table \ref{form1}.  In the table,
as a comparison, we have also shown
the value of $u_{+}(0)$ used in Ref. \cite{Cheng1}.
To give explicit $q^{2}$ dependent form factors, we fit our results
to the form
 \be
     F(q^2)=\frac{F(0)}{1-a(q^2/m_B^2)+b(q^2/m_B^2)^2}\,,
   \ee
 with the fitted ranges of $F_T(q^2)$ and $u_{\pm}(q^{2})$ being
  $0\leq q^2\leq 12 $  and
  $0\leq q^2\leq (m_B-m_{K_0^*})^2 $ GeV$^2$, respectively.
\begin{table}[h]
   \caption{Form factors at $q^{2}=0$ for the $B\to K_0^*$ transition}
   \label{form1}
   \begin{center}
  \begin{tabular}{|c|ccc|ccc|}
   \hline
   & \multicolumn{3}{c|}{this work} & \multicolumn{3}{c|}{\cite{Cheng1}} \\ \hline
         & $F(0)$ & $a$& $b$& $F(0)$& $a$& $b$ \\ \hline
   $u_+$ & $-0.26$ & $1.36$& $0.86$& $-0.26$& $1.52$& $0.64$ \\
   $u_-$ & $~0.21$ & $1.26$& $0.93$& $-$& $-$& $-$\\
   $F_T$ & $~0.34$ & $1.64$& $1.72$& $-$& $-$& $-$\\ \hline
   \end{tabular}
   \end{center}
\end{table}
In Fig. \ref{formfactors}, the form factors as
functions of $q^{2}$ are presented, where
(a) $u_{\pm}(q^2)$ and (b) $F_T(q^2)$.
In Table \ref{wcs}, we give
the  values of the relevant WCs at the scale of $\mu \sim 4.8$
GeV  \cite{CG-PRD}.
\begin{table}[h]
   \caption{Wilson coefficients for $m_t=170$ GeV and $\mu=4.8$ GeV.}
   \label{wcs}
   \begin{center}
   \begin{tabular}{|c|c|c|c|c|c|}
   \hline
  WC & $C_1$ & $C_2$ & $C_7$ & $C_9$ & $C_{10}$ \\ \hline
  & $-0.226$ & $ 1.096$ & $-0.305 $ & $4.186$ & $-4.559$
  \\ \hline
  \end{tabular}
   \end{center}
\end{table}
 With $|\lambda_{t}|\simeq 0.041$,
 we illustrate
the differential decay branching ratios for
$B\to K_0^* \nu \bar{\nu}$ and
$B\to K_0^* \ell^{+}\ell^{-}\ (\ell=\mu,\tau)$ as  functions of $s$
in Figs. \ref{Figbrnn} and \ref{Figbrll},
respectively.
 By integrating the differential ratios over $s=q^2/m^2_{B}$
for $B\rightarrow K_0^* \nu \bar{\nu }$
 and $B\rightarrow K_0^* \ell^{+}\ell^{-}$,
 we obtain
%
 \be
    Br( B\rightarrow K_0^* \nu \bar{\nu })&=&
        1.16\times 10^{-6}
\ee
and
\be
Br(B\to K_{0}^{*}e^{+}e^{-},K_{0}^{*}\mu^{+}\mu^{-},
K_{0}^{*}\tau^{+}\tau^{-})=
1.63\times 10^{-7}\,,\ 1.62\times 10^{-7}\,,\ 2.86\times 10^{-9}\,,
\ee
respectively. Note that the small branching ratio of the tau mode
is due to the highly suppressed phase space as shown
in Fig. \ref{Figbrll}.
 We also present
the longitudinal lepton polarization
asymmetries of $B\to K_0^* \ell^+\ell^-$ as  functions of $s$
in Fig. \ref{PLbrll}.
We note that our results for the electron mode are similar to those
for the muon one. As shown in Fig. \ref{PLbrll},
   $P_{L}(s)[B \to K^*_0 \mu ^{+}\mu ^{-}]$
  is close to $-1$  except those close to the end points of
  $q_{min}^{2}=4m_{\mu}^{2}$ and $q_{max}^{2}=(m_{B}-m_{K^{*}_{0}})^{2}$
  at which they are zero and $P_{L}(s)[B \to K^*_0\tau^{+}\tau^{-}]$
  ranges from $-0.5$ to 0, while the integrated values of
  $P_{L}$
  are $-0.97$, $-0.95$ and $-0.03$ for electron, muon and tau modes, respectively. It is clear that due to the efficiency
  for the detectability of the tau lepton, it is impossible to
  measure the tau lepton polarization in the near future.
  
Finally, we remark that our results are insensitive (sensitive)
to the value of $f_{B}$ ($f_{K^{*}_{0}}$).
For examples, 
 $Br(B\to K^{*}_{0}\ell\bar{\ell})$ ($\ell=\nu,e,\mu$) 
 decrease about $6\%$ by increasing $f_{B}=0.16$ to
 $0.20$ GeV, while they increase about $50\%$ by increasing
 $f_{K^{*}_{0}}=0.021$ to $0.025$ GeV.
%
\section{ Conclusions}
We have studied the exclusive rare decays of $B \to
K_{0}^*\ell\bar{\ell} $. We have calculated the form factors  for
the $B\to K_{0}^{*}$ transition matrix elements in the LFQM. We
have evaluated the decay branching ratios and the longitudinal
charged-lepton polarization asymmetries in the SM. Explicitly, we
have found that $Br(B \to K_{0}^*\ell\bar{\ell})\
(\ell=\nu,e,\mu,\tau)\ =(11.6,1.63,1.62,0.029)\times 10^{-7}$ and
the integrated longitudinal lepton polarization asymmetries of $B \to
K^*_0 \ell ^{+}\ell ^{-}\ (\ell=e,\mu,\tau)$
are $-0.97$, $-0.95$ and $-0.03$, respectively. It
is clear that some of the above p-wave B decays and asymmetries
can be measured at the ongoing as well as future B factories.

\section*{Acknowledgments}
This work is supported in part by the National Science Council of
R.O.C. under Contract
 \#s:NSC-95-2112-M-006-013-MY2,
  NSC-95-2112-M-007-059-MY3,
 NSC-95-2112-M-009-041-MY2, NSC-95-2112-M-277-001
 and NSC-96-2918-I-007-010.


\end{document}